\documentclass[preprint,12pt]{elsarticle}

\usepackage{amssymb}

\journal{arXiv}

\begin{document}

\begin{frontmatter}



\title{Bone Marrow Cytomorphology Cell Detection using InceptionResNetV2}


\author[inst1]{Raisa Fairooz Meem}

\affiliation[inst1]{organization={Department of Computer Science, American International University Bangladesh},
            addressline={Dhaka}, 
            city={Dhaka},
            country={Bangladesh}}

\author[inst2]{Khandaker Tabin Hasan}

\affiliation[inst2]{organization={Department of Computer Science, American International University Bangladesh},
            addressline={Dhaka}, 
            city={Dhaka},
            country={Bangladesh}, 
            email= { tabin@aiub.edu}}

\begin{abstract}
Critical clinical decision points in haematology are influenced by the requirement of bone marrow cytology for a haematological diagnosis. Bone marrow cytology, however, is restricted to reference facilities with expertise, and linked to inter-observer variability which requires a long time to process that could result in a delayed or inaccurate diagnosis, leaving an unmet need for cutting-edge supporting technologies. This paper presents a novel transfer learning model for Bone Marrow Cell Detection to provide a solution to all the difficulties faced for the task along with considerable accuracy. The proposed model achieved 96.19\% accuracy which can be used in the future for analysis of other medical images in this domain.
\end{abstract}

\begin{graphicalabstract}
\includegraphics{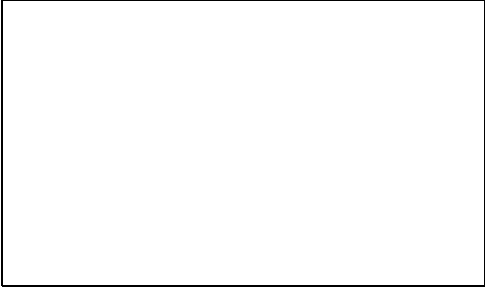}
\end{graphicalabstract}

\begin{highlights}
\item Explore the application of Transfer Learning for Bone Marrow Cell Detection
\end{highlights}

\begin{keyword}
Transfer Learning \sep Bone Marrow Cell Detection \sep InceptionResNet
\PACS 0000 \sep 1111
\MSC 0000 \sep 1111
\end{keyword}

\end{frontmatter}


\section{Introduction}
A haematological diagnosis is mainly based on a bone marrow analysis \cite{Yu2019}. It is carried out to investigate a haematological problem that is clinically suspected, as part of lymphoma staging procedures, and to evaluate the bone marrow response to chemotherapy in acute leukaemias. It is also examined in myelodysplastic syndrome (MDS) and myeloproliferative neoplasm (MPN) to detect several factors such as dysplastic cellular morphology, blast excess, and cellularity of the subject etc. \cite{Bruck2021}. A hematopathologist will gather data from the various parts of a bone marrow investigation and combine it with clinical data to arrive at a final diagnostic interpretation. There are currently no clinical-grade solutions for bone marrow cytology, even though a number of commercial computational pathology workflow support systems have been created for the analysis of peripheral blood cytology. The reliability of a disease diagnosis depends on the accuracy of identification of histopathological image slides for which, a computerized process, namely machine learning can be an affordable yet efficient solution, replacing the dependency on human experts \cite{Fan2021}. Bone marrow aspirates are more sophisticated cytological specimens than blood film cytology. Aspirates have a limited number of cytologically useful regions, a large amount of non-cellular debris, and a wide variety of cell types that are frequently aggregated or overlapping. This has made computational pathology problems related to bone cytology somewhat difficult \cite{Tayebi2022}.

Malignant and nonmalignant illnesses affecting the haematopoietic system can be diagnosed by examining and differentiating the cell morphologies of the bone marrow (BM). Despite the availability of numerous advanced techniques, including cytogenetics, immunophenotyping, and increasingly molecular genetics, cytomorphologic evaluation is still a crucial initial step in the diagnosis of many intra- and extramedullary diseases. Since the process has proven challenging to automate, in a clinical workflow, human experts are still primarily responsible for microscopic analysis and single-cell morphological classification. Manual evaluation of BM smears, however, can be laborious and time-consuming and is greatly reliant on the knowledge and experience of the examiner, particularly in cases when the results are unclear. As a result, the availability and expertise of qualified experts is a constraint on the number of high-quality cytological examinations, although examiner classifications have been found to be vulnerable to significant inter- and intrarater variability. Furthermore, the procedure is challenging to combine with other diagnostic techniques that provide more quantitative data because the analysis of individual cell morphologies is fundamentally qualitative \cite{Matek2021}.

Deep learning techniques for object detection include regional CNN (R-CNN), and fast and Faster R-CNN. These methods rely on region proposals for object recognition, followed by a different technique like object classification, which makes them difficult to train and ineffective computationally \cite{Tayebi2022}. Deep learning models, in particular convolutional neural networks (CNN), are extensively employed for picture categorization tasks \cite{Muj2022}. But the main obstacle to widely applying it for the medical image analysis sector is the unavailability of large datasets as small datasets for training lead to overfitting results or convergence \cite{Sw2019}. Assigning individual bone marrow cells or non-cellular objects to one of the multiple discrete classes based on sophisticated and nuanced cytological traits presents another challenge in object categorization. In MDS, when morphological dysplasia causes subtle cytological alterations, this complexity rises. Many tried to solve the issues by fine-tuning the Faster R-CNN. This method, however, proved to be inefficient operationally and is probably not adaptable to a clinical diagnostic workflow. To enable bone marrow aspirate cytology, unique, effective, and scalable computational pathology methodologies are required. Approaches that add full end-to-end automation, that is, from raw images to bone marrow cell counts and classification \cite{Tayebi2022}.

Transfer learning (TL) gained the attention of researchers working on medical image analysis in recent years. Transfer learning is a branch of CNN that has had its weights initialized before and can be trained quickly and accurately due to being trained on a bigger dataset. The main advantage of using transfer learning is that instead of starting from scratch with the random initialization, the weights used can be applied to categorize another entirely different dataset \cite{Kan2021}. Most commonly used models are often trained on the ImageNet dataset, a massive collection of images created for use in the development of visual object recognition software. The project has manually annotated over 14 million photographs to identify the things they depict, and at least one million of the images also include bounding boxes. Although the actual images are not ImageNet's property, the database of annotations for third-party image URLs is freely accessible. A deep CNN model that has previously been trained on a huge dataset is used for the procedure. The CNN model is further trained (fine-tuned) using a new dataset that has fewer training images than the previously trained datasets. While the latter layers of CNN models concentrate on more abstract features, the first layers of CNN models frequently learn features like edges, curves, corners, etc. Frequently, the remaining layers are relocated to carry out a new classification while the completely linked layer, SoftMax layer, and classification output layer are discharged \cite{Den2018}. Transfer Learning was a common option when the available dataset was small since it did not require huge datasets to train the model with randomly initialized weights from start \cite{Faria2018}.

Inception and ResNet, two of the most effective models, were combined to create the well-known TL model known as ResNet. This model, which was developed using the ImageNet database, substitutes batch normalization for convolutional layer summation. To prevent overfitting, input units are set to 0 at random in the dropout layers during the training phase. In order to use a one-dimensional data array in the subsequent layers, the model additionally employs a flattering strategy. The same procedure is used to generate a feature vector from the output data. By connecting to the classification model's final layer, the model creates a fully connected layer with a batch size of 32 and a "binary cross-entropy" loss function \cite{Faruk2021}.

In this study, a small dataset of bone marrow smears is analyzed using InceptionResNetV2 to find the cells suggestive of different haematological illnesses. This article is divided into five sections: the second section highlights the relevant improvements made in this field to date; the third describes the research's methods and supplies; the fourth examines the findings; and the fifth closes with recommendations for further research.

\section{Literature Review}
Recently, Convolutional Neural Networks (CNN) and Deep Learning (DL) are being utilized extensively for medical image analysis. Naturally, the use of Transfer Learning (TL) is also promoted in this field, given that it provides a workaround for the need for huge training datasets.

To offer the most recent status of this subject, Morid, Mohammad Amin et al. \cite{Morid2021} reviewed selected articles published after 2018 on the utilization of different ImageNet-trained transfer learning models for the analysis of clinical images. According to their analysis, the use of models depends on the type of images that are being processed. For X-ray, endoscopic, or ultrasound pictures, for instance, Inception models are often utilized, whereas VGGNet performs comparably better for OCT or skin lesion images. No matter the type of image or organ, the top 4 models are Inception, VGGNet, AlexNet, and ResNet in order of most frequent usage. Only 3 papers mentioned the InceptionResNet model because it is a relatively novel method. 

The literature assessment undertaken by Kim, Hee E., et al. \cite{Kim2022} also supported the choice of Inception and ResNet models. They assessed 425 articles on medical image classification using transfer learning models that were published till 2020, and the result indicated the usefulness of these models considering the scarcity of data. Their study on reducing costs and time for clinical image detection provided some interesting outputs such as: 1. for most transfer learning models, fine-tuning only the last fully connected layers provides a better result than starting from scratch. 2. Unfreezing the convolutional layers while maintaining a low learning rate from top to bottom manner can improve the overall performance of the model. According to their study on various models used in the selected articles, Inception and ResNet models were proven the most effective feature extractors which are able to provide impressive accuracy even after reducing the time required and computational costs.

Faruk, Omar et al. \cite{Faruk2021} applied 4 different transfer learning models, namely Xception, InceptionV3, InceptionResNetV2, and MobileNetV2 for tuberculosis detection using X-ray images. Each model consists of three MaxPooling2D levels, four Conv2D layers, a flattened layer, two dense layers, and a ReLU activation function. With a few tweaks in the final layers, the thickest and last layer, SoftMax, functions as an activation layer. Layers like average pooling, flattening, dense, and dropout are used to customize the results. InceptionResNetV2 outperformed the other 4 models in terms of accuracy (99.36\%).

Matek, Christian, et al. \cite{Matek2021} proposed a model named ResNeXt to classify Bone Marrow Cell Morphology. ResNeXt replicates a structural component that combines several transformations with the same topology. In addition to the depth and width dimensions, it also highlights cardinality (the size of the set of transformations) as a new dimension in comparison to ResNet. They preferred this particular model as it was previously applied for categorizing peripheral blood smears. They selected only 32 cardinality hypermeters for the research. They used a single-centre strategy, preparing all the BM smears included for training in the same laboratory and digitizing them using the same scanning tools. The network reported in this study performs in a highly encouraging manner where the external validation shows that the strategy is generalizable to data gathered in various settings, albeit being difficult due to the small amount of available data.

Yu, Ta-Chuan, et al. \cite{Yu2019} proposed a deep CNN architecture to automatically identify and classify Bone Marrow cells. The dataset consisted of Liu’s stained images of bone marrow smears of patients at the National Taiwan University Hospital. The proposed model considered and applied several factors such as group normalization, colour shift, Gaussian blur etc. for various tasks such as model training or augmentation of data and achieved 93.6\% accuracy.

Rahman, Jeba Fairooz, and Mohiuddin Ahmad \cite{Kuet2022} conducted a comparative study on 4 transfer learning models, namely AlexNet, VGG-16, ResNet50, and DenseNet161 to detect Acute Myeloid Leukaemia (AML) that is mainly distinguished by the traces of immature leukocytes in blood and bone marrow. They modified the applied models using a binary classifier and ReLu for activation as they aimed to detect mature and immature leukocytes only. Among the models, AlexNet achieved an accuracy of 96.52\% for detecting leukocytes.

Many of the research works on haematological disease analysis focus on Acute Lymphoblastic Leukemia (ALL) Cell detection.  It is a type of cancer that damages organs and tissues involved in blood production and the blood circulation system as well. It damages the production of healthy white blood cells that are responsible for protecting the body from various diseases \cite{Af2021}.

Kumar, Deepika, et al. \cite{KD2020} proposed a Dense CNN model for accurate identification of ALL and Multiple Myeloma and compared the performance with multiple machine learning methods, namely, Support Vector Machine (SVM), VGG16, Naïve Bayes, Random Forest (RF), Decision Tree (DT) etc.  Adam Optimizer was used to train the model consisting of multiple Convolution, Pooling, and Connected layers with Sigmoid loss function. The model achieved an accuracy of 97.25\%.

Liu, Ying, et al. \cite{LY2019} and Ramaneswaran, S, et al. \cite{Ram2021} both applied transfer learning models for classifying ALL Cells. Liu, Yin, et al. applied InceptionResNetV2 as a backbone network to identify lymphoblastic leukaemia cells using 2 stages deep bagging ensemble learning technique while Ramaneswaran, S, et al. used InceptionV3 as the feature extractor with XGBoost Classifier instead of SoftMax. Liu, Ying, et al. trained the model using the bagging ensemble learning to minimize data imbalance and produced two separate models that were later combined to enhance the final performance. The model achieved an F1 score of 0.88 for classifying the cells. The hybrid InceptionV3 with XGBoost Classifier model proposed by Ramaneswaran, S, et al. achieved an accuracy of 97.9\%.

Abir, Wahidul Hasan et al. \cite{Abir2022} conducted a comparative study on the performance of several transfer learning models for identifying ALL cells. The models were InceptionV3, InceptionResNetV2, ResNet101V2, and VGG19. They applied Local Interpretable Model-agnostic Explanation (LIME) algorithm and XAI to ensure the reliability and validity of the proposed model. They applied a stratified k-fold cross-validation technique to remove the imbalance in the dataset for class distribution. InceptionV3 model achieved 96.65\% accuracy for the validation set, outperforming the other models, though for the training set, InceptionResNetV2 had the highest accuracy (99.14\%).

Tayebi, Rohollah Moosavi, et al. \cite{Tayebi2022} proposed a You-Only-Look-Once (YOLO) model for automated bone marrow cytology. An end-to-end system based on deep learning was created for the experiment. The model automatically detects regions that are suitable for cytology from a digital whole slide image of a bone marrow aspirate, and then it identifies and categorizes every bone marrow cell in each region. This comprehensive cytomorphological data is represented by a cytological patient fingerprint known as the Histogram of Cell Types (HCT), which quantifies the probability distribution of bone marrow cell classes. The proposed system provided an impressive accuracy considering region detection (0.97 accuracies and 0.99 ROC AUC), cell detection and cell classification (0.75 mean average precision, 0.78 average F1-score, and a log-average miss rate of 0.31).

Loey, Mohamed, Mukdad Naman, and Hala Zayed \cite{Loey2020} claimed to achieve 100\% accuracy in leukaemia detection using transfer learning models. They applied 2 models for the task. The first model used AlexNet for feature extraction and several classifiers, namely SVM (Linear, Gaussian, and Cubic), DT, Linear Discriminants (LD) and K-NN for the classification process. The model with SVM-Cubic achieved the highest accuracy among other classifiers (99.79\%). The 2nd model used AlexNet both for feature extraction and classification. This model achieved 100\% accuracy in detecting leukaemia using blood cell image slides.

In conclusion, the field of brain and lung illness detection, particularly different types of cancer, uses TL models most frequently. However, because of how well they worked, researchers started using them to find problems in other organs as well. We are motivated to conduct more studies in this area because several models have demonstrated outstanding accuracy in the identification of haematological diseases from Bone Marrow Cell images.

\section{Methodology}
\subsection{Dataset}
The dataset from the work of Matek et al, \cite{Mat2021} consists of 171,375 cells from bone marrow smears of 945 patients stained with the May-Grünwald-Giemsa/Pappenheim stain that has been de-identified and expertly annotated. The cohort's diagnosis distribution includes a range of haematological conditions consistent with the sample entry of a sizable laboratory with a focus on leukaemia diagnostics. A brightfield microscope with a 40x magnification and oil immersion was used to get the images. 20\% of the dataset, chosen at random, was used for the thesis. The images are divided into 21 categories based on the haematological disease or aspect they present. The categories are mentioned in Table 1.

\begin{table}
    \centering
    \caption{List of Haematological diseases cell images of which are included in the dataset}
    \vspace{10pt}
\begin{tabular}{||c | c||} 
 \hline
 Abbreviation & Meaning \\ 
 \hline\hline
 ABE & Abnormal eosinophil \\ 
 \hline
 ART & Artefact \\ 
 \hline
  BAS & Basophil \\ 
 \hline
  BLA & Blast \\ 
\hline
  EBO & Erythroblast \\ 
 \hline
  EOS & Eosinophil \\
   \hline
 FGC & Faggott cell \\ 
 \hline
 HAC & Hairy cell \\ 
 \hline
 KSC & Smudge cell \\ 
 \hline
  LYI & Immature lymphocyte \\ 
 \hline
  LYT & Lymphocyte \\ 
\hline
  MMZ & Metamyelocyte \\ 
 \hline
  MON & Monocyte \\
   \hline
  MYB & Myelocyte \\ 
 \hline
  NGB & Band neutrophil \\ 
\hline
  NGS & Segmented neutrophil \\ 
 \hline
  NIF & Not identifiable \\
     \hline
  OTH & Other cells \\ 
 \hline
  PEB & Proerythroblast \\ 
\hline
  PLM & Plasma cell \\ 
 \hline
  PMO & Promyelocyte \\
  \hline
\end{tabular}
\end{table}

\subsection{Evaluation Criteria}
To evaluate the performance of the transfer learning model, a confusion matrix was utilized where four types of output were considered- True Positive (TP) where an output accurately identifies the existence of a condition, True Negative (TP) meaning an output correctly indicates the absence of a characteristic, False Positive (FP) where the existence of a condition is falsely indicated and False Negative (FN) where the output wrongly identifies the existence of a condition. The concept of the confusion matrix is further explained in Table 2:

\begin{table}
    \centering
    \caption{Explanation of TP, TN, FP, and FN}
    \vspace{10pt}
\begin{tabular}{||c | c | c||} 
 \hline
 I/O & Output Negative & Output Positive \\ [0.5ex] 
 \hline\hline
 Input Positive & False Positive (FP) & True Positive (TP) \\ 
 \hline
 Input Negative & True Negative (TN) & False Negative (FN) \\ [1ex] 
 \hline
\end{tabular}
\end{table}

Five criteria are used to evaluate the output: accuracy, loss, precision, recall, and AUC. 

The evaluation of how closely the expected result matches the actual result is called Accuracy, expressed as a percentage. It is computed by dividing the sum of true positive and true adverse events by the total number of outcomes that could occur:

Accuracy = (TP + FN) / (TP + TF + FP + FN)

The degree to which predicted values accord with one another is known as Precision. One can determine the true positive by dividing the true positive by the sum of true and false positives.

Precision = TP / (TP + FP)

The gross number of true positives is divided by the total number of true positives and false negatives to determine the Recall value of an experiment.

Recall = TP / (TP + FN)

For this experiment, ‘32’ was selected as the batch size along with only 5 epochs. Google Colaboraty was used as the experiment's coding platform, which also utilized GPU runtime. ‘Categorical Loss Entropy’ was selected for the experiment. According to Zaheer et al \cite{Za2019}, the Adam optimizer, derived from 'Adaptive Moment' outperforms all the other optimizers in terms of accuracy which is why it was selected for this experiment.

\section{Output Evaluation}

\begin{table}
    \centering
    \caption{Evaluation of InceptionResNetV2 for Bone Marrow Cell Detection}
\begin{tabular}{||c | c| c | c | c | c ||} 
 \hline
 Set & Loss & Accuracy & Precision & Recall & AUC \\  
 \hline\hline
 
 Training & 5.7916 & 96.39\% & 0.6214 & .6171 & 0.8472 \\ 
 \hline
  Validation & 7.2734 & 96.19\% & 0.6 & 0.5968 & 0.8297 \\ 
  \hline
\end{tabular}
\end{table}

The result points out several things- first, the difference between the accuracy of training and validation sets is very little (0.2\%). The loss values are quite high for both sets, even more in the case of the validation set (7.2734). Precision and recall values for both sets are almost similar, even with each other.

\section{Conclusion}
Transfer learning models reduce the possibility of false detection along with the complexity, human participation requirements, required time etc. which is beneficial for biological image segmentation and can ensure that everyone, including those with minimal resources, receives competent medical care. The paper examines the effectiveness of the InceptionResNetV2 model for bone marrow cell detection, a crucial component of the analysis of blood disorders that have not received enough attention from researchers who are curious about the potential uses of transfer learning models in this field. To better understand them, other criteria were applied to the output the model produced. The chosen model had a 96.19\% accuracy rate. This work points to potential areas for further research, such as: 

\begin{itemize}
\item Comparing several transfer learning models to see whether any model can outperform the InceptionResNetV2 models.
\item A follow-up investigation to improve InceptionResNetV2's Bone Marrow Cell Detection accuracy.
\item Other bone diseases and illnesses in other organs can be detected using the InceotionResNetV2 model.
\end{itemize}

\end{document}